# Engineering Thermal and Electrical Interface Properties of Phase Change Memory with Monolayer MoS$_2$

Christopher M. Neumann,[1] Kye L. Okabe,[1] Eilam Yalon,[1,2] Ryan W. Grady,[1] H.-S. Philip Wong,[1] and Eric Pop[1,3,a)]

[1]*Dept. of Electrical Engineering, Stanford University, Stanford, California 94305, USA*
[2]*Dept. of Electrical Engineering, Technion – Israel Institute of Technology, Haifa 32000, Israel*
[3]*Dept. of Materials Science and Engineering, Stanford University, Stanford, California 94305, USA*

## Abstract

Phase change memory (PCM) is an emerging data storage technology, however its programming is thermal in nature and typically not energy-efficient. Here we reduce the switching power of PCM through the combined approaches of filamentary contacts and thermal confinement. The filamentary contact is formed through an oxidized TiN layer on the bottom electrode, and thermal confinement is achieved using a monolayer semiconductor interface, three-atom thick MoS$_2$. The former reduces the switching volume of the phase change material and yields a 70% reduction in reset current versus typical 150 nm diameter mushroom cells. The enhanced thermal confinement achieved with the ultra-thin (~6 Å) MoS$_2$ yields an additional 30% reduction in switching current and power. We also use detailed simulations to show that further tailoring the electrical and thermal interfaces of such PCM cells toward their fundamental limits could lead up to a six-fold benefit in power efficiency.

a)Author to whom correspondence should be addressed: epop@stanford.edu

Phase change memory (PCM) is an emerging storage class memory technology, wherein the typical cell consists of a chalcogenide-based phase change material (commonly $Ge_2Sb_2Te_5$) contacted by top and bottom electrodes (TE and BE, respectively). The metal electrodes are used to apply voltage or current pulses to the phase change material, inducing its (reversible) transformation between amorphous and crystalline phases.[1] The phases possess electrical resistivity differing by up to four orders of magnitude, which can be read out to store binary or analog logic states (i.e. by gradual or partial programming) for neuromorphic applications.[2,3]

While PCM is already being used in computing systems, concerns remain over its relatively high reset current and power. To mitigate this problem, two approaches have often been taken: (1) reducing the volume of phase change material, or (2) improving the thermal confinement of the cell.[1] Reducing the switching volume entails scaling the contact area, either lithographically or by using nanoscale electrodes such as oxide filaments,[4-6] carbon nanotubes,[7-9] or graphene edge contacts.[10] Improving the thermal confinement requires trapping Joule heat by confining current flow through the phase change material,[11] using more thermally resistive materials,[12,13] or adding interfacial layers to the electrode contacts.[14-17] In earlier experiments, two-dimensional (2D) materials, like graphene, have already been used as an interfacial layer.[14,17] However, the high thermal and electrical conductivity of graphene counteracts our attempts to confine heating.[14]

In this work, we combine enhancements from both categories above to demonstrate power-efficient PCM cells. To improve thermal confinement, we use monolayer molybdenum disulfide ($MoS_2$) grown by large-scale chemical vapor deposition (CVD).[18] Inserting this three-atom-thick (~6.15 Å) layer at the interface between the BE and the phase change material limits heat loss through this interface. To reduce the contact area, we form a narrow metal-oxide filament in the thin oxide on top of the TiN BE.[4-6] We also use finite element modeling to gain additional insight into the benefits of further optimizing such interface modifications, toward fundamental physical limits, for power-efficient PCM.

We fabricate filamentary BE PCM with a $MoS_2$ interfacial layer [Fig. 1(a)], as well as three types of control devices: (1) filamentary BE PCM with graphene as an interfacial layer, (2) filamentary BE PCM without a 2D material, and (3) conventional PCM of the same dimensions without a filament nor a 2D layer. For all device types, we start with planarized TiN BEs which are ~150 nm in diameter. For filamentary devices, the BEs were first cleaned through an Ar sputtering process before being exposed to water for at least one hour to ensure that a thin oxide layer forms. For devices with 2D materials, we use polymer-assisted wet transfer techniques[19,20] to place the 2D material layer on the BE substrate and pattern it using e-beam lithography. The patterned area covers the entire BE and ~25 nm past the BE edges to account for overlay placement margin. Next, we DC sputter and lift-off 30 nm of $Ge_2Sb_2Te_5$ (GST) capped *in situ* with 20 nm of TiN. We previously showed that such GST depositions can be done with minimal damage to graphene.[14] Figure 1(b) shows Raman spectroscopy data with a strong $MoS_2$ signal both before and after GST deposition, confirming that $MoS_2$ can also withstand this process and Fig. 1(c-d) displays transmission electron microscopy (TEM) cross-sections showing the interfacial $MoS_2$ layer. For devices without filaments, the GST/TiN layer is deposited *in situ* after the Ar sputter clean to prevent any native oxide formation. After GST/TiN lift-off, we pattern and lift-off an additional 20 nm of sputtered TiN and 40 nm of Pt to form the probe pads and top electrode. Finally, devices are annealed in air at 180°C for at least one hour to crystallize the GST layer into the fcc phase.

In samples with oxidized BEs, we must first form a filament before we see memory operation. This is done by applying 1/50/1 ns rise/width/fall pulses of increasing bias until the filament is formed and we observe >200 μA of current during the pulse. In more than 50 measured devices, this occurs between 1.5 to 2 V. To demonstrate the filamentary nature of these oxidized BEs, we measured devices of various BE sizes with and without oxidation. Figure 2(a) shows that the current required to reset devices without oxide increases with increasing BE area, as is expected for a PCM device. However, in Fig. 2(b), devices with the oxide layer show no dependence on BE area and exhibit significantly lower reset currents. Given this trend, we conclude that in devices with an oxide layer, the effective BE size is related to the filament's size rather than the physical BE size.

We measure the resistance of all devices with a 50 mV DC bias. For set and reset programming, we use 1/50/2000 ns and 1/50/1 ns rise/width/fall pulses, respectively. The general layout of the measurement system and endurance information can be found in the supplementary material. Next, we compare the reset current of the $MoS_2$ filament-contacted devices with the control devices. To do this we first cycle all measured devices at least 1000 times to ensure consistent and reliable operation. Subsequently, cells are set into the low resistance state. We apply a series of increasing amplitude reset pulses to the device and capture the transient current through an oscilloscope. To obtain the current, we measure the voltage across the 50 Ω input of the oscilloscope. We note that, where a pulse current is given, we refer to the peak current rather than the average. For peak power calculations, we assume nominal applied voltage and peak current.

In Fig. 3, we plot the peak current, nominal voltage, and peak power required to reset devices of each type. Comparing filament-contacted devices with and without the $MoS_2$ layer, we see about a 30% reduction in the current, but the reset voltage remains constant. We attribute this reduction to the additional electrical and thermal resistance at the BE-GST interface. Conversely, for the graphene-interfaced PCM, we see a significant increase in the switching current. This is similar to observations in prior work[14] where a carefully patterned graphene layer gave a current reduction, but a graphene area much larger than the BE led to larger programming current. In our devices, the graphene is patterned to be slightly larger than the BE, but the filament's effective area is much smaller. Because graphene has relatively high in-plane thermal and electrical conductivity, it acts as a "spreader" for the heat and current. However, as shown in Fig. 4, $MoS_2$ has significantly lower (~20×) in-plane thermal conductivity than graphene. In addition, the electrical conductivity of undoped $MoS_2$ is orders of magnitude lower than graphene and this 2D layer can effectively be considered an insulator.[21,22] Because of this, cell power-efficiency can be improved even when $MoS_2$ is not precisely patterned.

We performed finite-element simulations to examine how much of an impact we can expect by modifying the thermal and electrical interfaces of a typical mushroom cell device. The key is to determine the expected power benefits of adding a 2D material layer and also to understand the "ultimate limits" of a perfect interface material with ideal thermal and electrical properties. We assume a device structure similar to our fabricated design with 100 nm diameter TiN BE and 30 nm thick GST. The top metal stack is identical with 40 nm of TiN and Pt. To simplify the simulation, we chose not to include the filamentary structure, and the 2D material layer is modeled as a simple increase in the electrical and thermal resistance of the interface. We take into account the temperature-dependent electrical contact resistance, $\rho_c(T) = \rho(T)t_{eff}$, by tying it to an effective GST thickness, $t_{eff}$, where the electrical resistivity of GST, $\rho(T)$, is obtained from Ref. 23. In other words, an effective GST thickness of 10 nm would give an interface that has the same electrical resistance as 10 nm of GST at that temperature. The temperature dependence of the TBR is not taken into account as it is expected to be negligible for $MoS_2$[24] above room temperature.

In Fig. 5, we show simulated cross-sectional temperature profiles of the device at the end of a 1.2 V, 50 ns pulse. For a given voltage, we note that while increasing the thermal resistance of the interface results in a higher temperature, increasing the electrical resistance reduces the temperature due to the lower current. It is important to note that, for simplicity, we are assuming that the PCM is the dominant electrical resistance in our "circuit." However, if we increase both boundary resistances, we see that the temperature is similar. In Fig. 6(a-b), we plot the maximum temperature change and current against a wide range of GST-BE interface parameters. We then calculate the relative power efficiency of the cell by normalizing the temperature increase with the power expended, as shown in Fig. 6(c). From these results, we see that increasing the thermal or electrical resistance can provide a 2× improvement in efficiency, and maximizing both shows a 6× benefit.

While it may be expected that thermal and electrical resistance should be maximized to increase power efficiency, there are both consequences and limits to doing so. Large increases in electrical resistivity will increase the switching voltage (requiring higher voltage transistors and introducing additional array-level power consumption) and decrease the memory window (limiting multi-level cell designs). In the case of increased thermal resistivity, In the case of increased thermal resistivity, we need to consider two possible trade-offs: the effect that thermal confinement has on the quenching process and the finite TBR of a single 2D material interface. For our mushroom cell, it is impractical to reach the quenching limit[25] as heat will dissipate through the TE and bulk GST even if the BE interface is thermally resistive (details can be found in the supplementary information section S3). While a thicker layer could be engineered to have very high thermal resistance, some heat generated in this layer would be trapped further from the phase change region. Therefore, we expect that using layered 2D heterostructures (e.g. a $MoS_2/WTe_2$ bilayer) will further increase the TBR while preserving the atomic thinness of the interface. These van der Waals heterostructures have been shown to have highly tunable properties[26] in addition to high TBR which exceeds that of bulk material interfaces.[27] However, future work must consider the effect of the relatively high (when compared to bulk insulators) in-plane thermal conductivity of these materials as well.

In conclusion, we fabricated a PCM device which combines the area reduction of a metal-oxide filament with the thermal confinement of monolayer $MoS_2$ at the BE-GST interface to reduce the switching current by 70% and 30%, respectively. Finite element simulations illustrate that increasing electrical or thermal boundary resistances individually can improve the power efficiency of typical PCM by more than 2×, and by 6× if both are increased up to near the fundamental limits. However, reaching these enhanced thermal boundary resistances will require novel approaches, such as the use of 2D heterostructures at the interface.

**Figures**

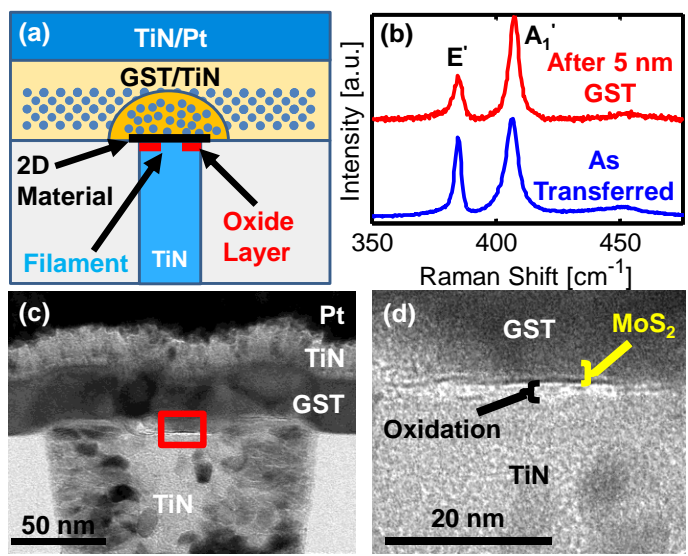

FIG. 1. (a) Cross-section schematic showing the position of the oxide filament and 2D material layers in a PCM device. (b) $MoS_2$ Raman spectra before (bottom, blue) and after (top, red) 5 nm deposition of GST, probed with 532 nm laser. Strong signal after GST sputtering indicates $MoS_2$ remains present. The $A_1'$ peak shows little change of position (405.9 cm$^{-1}$ before, 406.3 cm$^{-1}$ after) and FWHM (6.0 cm$^{-1}$ before, 5.8 cm$^{-1}$ after). The E$'$ peak position is also unchanged (384.9 cm$^{-1}$ before, 385.0 cm$^{-1}$ after), but the FWHM is broadened (3.5 cm$^{-1}$ before, 5.2 cm$^{-1}$ after). Peak center and full width half maximum (FWHM) values are extracted from peak fits. (c) TEM cross-sections of a PCM device fabricated for this work, with the $MoS_2$ interfacial layer. BE diameter is ~150 nm. (d) Zoomed in image of the red boxed area in (c) showing a ~2 nm oxidation layer on the TiN, just below the monolayer of $MoS_2$.

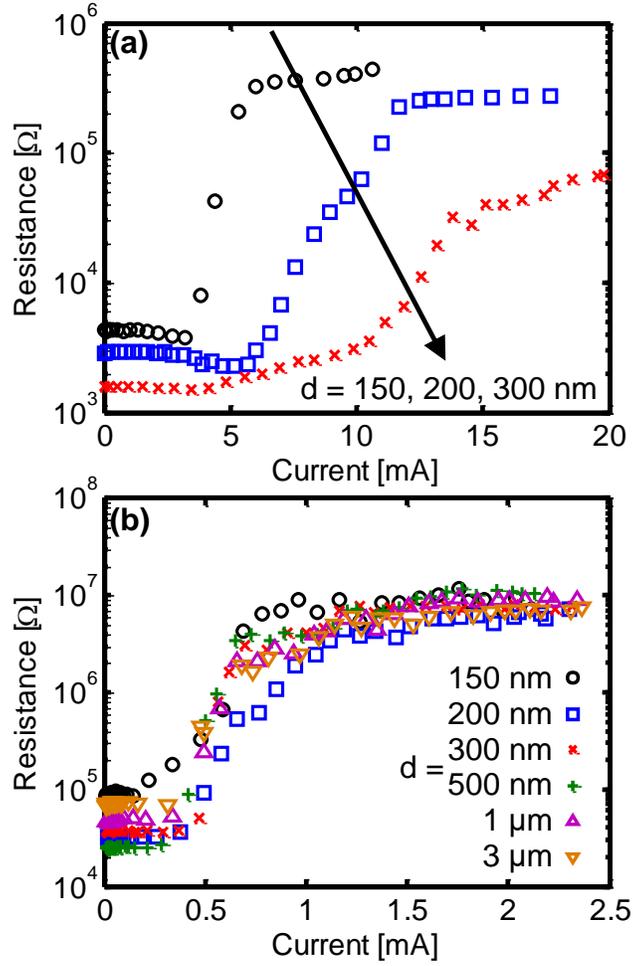

FIG. 2. DC read resistance vs. pulse current magnitude for various nominal BE diameters, *d*, without (a) and with (b) thin oxide layers on the BE. Cells without the oxide layer show expected scaling behavior, whereas those with oxidized BE have no dependence on BE diameter, indicating filamentary conduction through the oxide.

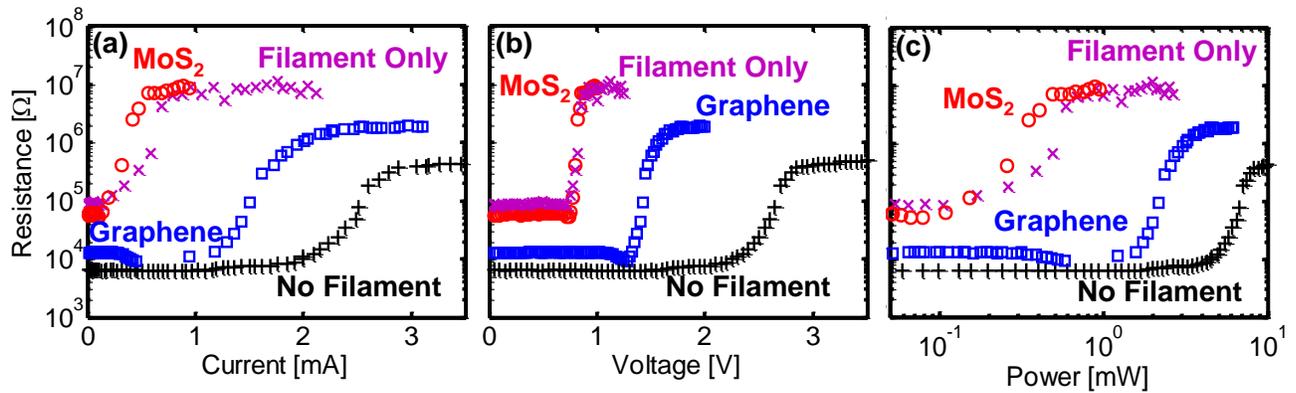

FIG. 3. DC read resistance vs. (a) current, (b) voltage, and (c) power during reset for the PCM with MoS$_2$ interfacial layer and oxide filament (red circle) and control devices: oxide filament only (purple x), graphene interfacial layer and oxide filament (blue square), and without oxide filament (black cross). Note that devices with the MoS$_2$ interfacial layer show more than 30% reduction in switching current and power compared to devices with filament only.

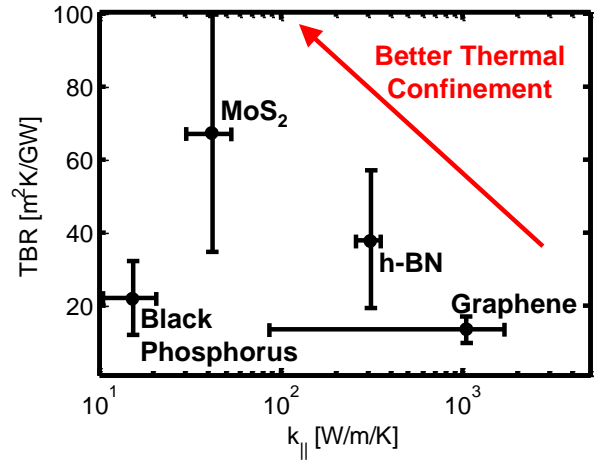

FIG. 4. Comparing thermal properties of some common 2D materials, where experimental data are available: range of measured thermal boundary resistance (TBR) vs. in-plane thermal conductivity ($k_{\parallel}$),[29-37] near room temperature. The lower end of the $k_{\parallel}$ range corresponds to confined samples (i.e. graphene nanoribbons)[37] or to samples with higher defect density. Overall, better thermal confinement for PCM could be achieved using 2D material interfaces with lower $k_{//}$ and higher TBR, e.g. 2D materials with heavier atomic masses (like $WTe_2$)[38] or with 2D material heterostructures.[26,27] At the higher PCM operating temperatures, the TBR and $k_{\parallel}$ will likely be lower due to increased phonon population and scattering, respectively.

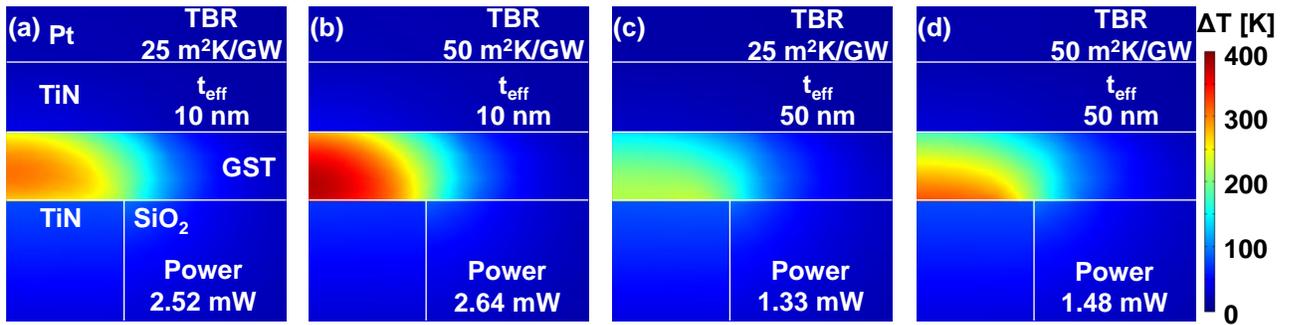

FIG. 5. Simulated temperature rise ($\Delta T$) for a 100 nm BE after 50 ns of 1.2 V bias for devices with (a) nominal boundary resistances as listed, (b) increased thermal boundary resistance (TBR), (c) increased electrical boundary resistance (as $t_{eff}$ of GST), and (d) increased both electrical and thermal boundary resistances. The values of TBR and effective thickness of electrical resistance ($t_{eff}$) are listed in each panel. Increasing TBR improves thermal confinement in the phase change layer, while increasing the electrical contact resistance raises the power density at the interface and generates more heat per input power.

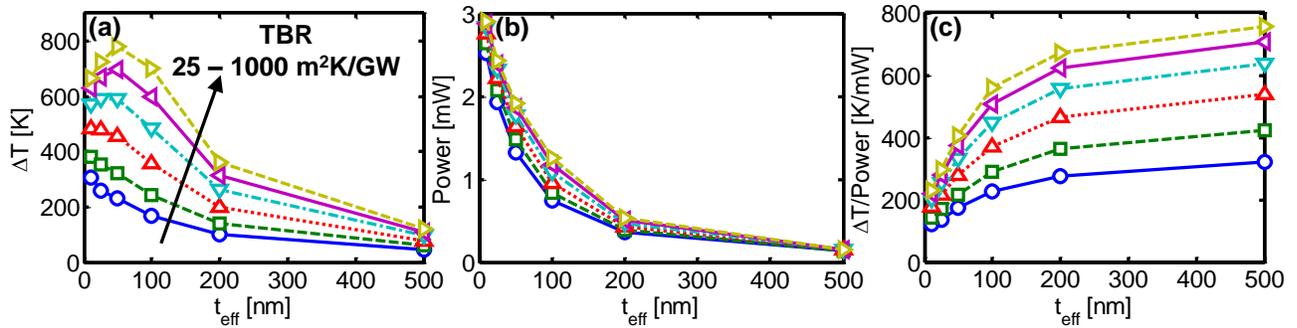

FIG. 6. (a) Maximum temperature rise and (b) power vs. effective thickness of electrical resistance ($t_{eff}$) for a 100 nm wide BE with varying thermal boundary resistance (TBR) under identical bias conditions (1.2 V for 50 ns). (c) Temperature rise per input power, corresponding to the thermal resistance of the device. Increasing both electrical and thermal boundary resistance improves the heating efficiency (as $\Delta T/P$) of such devices.

## Acknowledgements

Work was performed at the Stanford Nanofabrication Facility (SNF) and Stanford Nano Shared Facilities (SNSF), supported by the National Science Foundation (NSF) as part of the NNCI under award 1542152. This work was supported by member companies of the Stanford Non-volatile Memory Technology Research Initiative (NMTRI) and by the NSF EFRI 2-DARE grant 1542883. K.L.O. acknowledges funding from the Semiconductor Research Corporation (SRC) task 2826. RWG acknowledges support from the NSF Graduate Research Fellowship under grant DGE-1656518.

## Supplementary Material

See supplementary material for additional information on the device measurement setup and transient response.